\documentstyle[amsfonts,12pt,thmsa,sw20aip]{article}

\input tcilatex
\begin{document}

\author{F.P. Zen\thanks{%
Address after August 1$^{st}$, 1999 : Institute of Advanced Studies,
Research School of Physical Sciences and Engineering, Australian National
University, Canberra ACT 0200, Australia.}, B.E. Gunara\thanks{%
email : bobby@fi.itb.ac.id}, R. Muhamad\thanks{%
email : roby17@indosat.net.id}, and D. P. Hutasoit\thanks{%
email : davidphu@flashmail.com} \and Theoretical High Energy Physics Group,
\and Theoretical Physics Laboratory, \and Department of Physics, Bandung
Institute of Technology, \and Jl. Ganesha 10 Bandung, 40132, INDONESIA.}
\title{BPS States and Vacuum Manifold of $SU_q(n)$ Georgi-Glashow Model}
\date{THEP-PHYS-ITB-5-99\\
July 1999}
\maketitle

\begin{abstract}
We construct the Georgi-Glashow Lagrangian for gauge group $SU_q(n)$.
Breaking this symmetry spontaneously gives $q$-dependent masses of gauge
field and vacuum manifold. It turned out that the vacuum manifold is
parameterized by the non-commutative quantities. We showed that the monopole
solutions exist in this model, which is indicated by the presence of the BPS
states.
\end{abstract}

\section{Introduction}

The notion of the Lie group has been generalized by Drinfel'd[1], Jimbo[2],
and Woronowicz[3]. Their generalized Lie group, i.e., noncommutative and
non-cocommutative Hopf algebra, is now known as the quantum group under an
enthusiastic study by lot of mathematicians and physicists. Several authors
have attempted to quantize or $q$-deform the Lorentz group[4].

On the other hand, the Georgi-Glashow model, which is a simple theory in 3+1
dimensional, has been studied by many authors such as 't Hooft-Polyakov[5]
and Julia-Zee[6] . This model has a solitonic solution, called monopole, in
the Higgs vacuum[5, 6].

The purpose of this paper is to generalize the Georgi-Glashow model for the
case of quantum group, and to show that there exists a solitonic solution in
general case. We shall be concerned only quantum group $SU_q(n)$ with the
simplest example of quantum group, $SU_q(2)$, since it reduces to $SU(2)$
for $q=1$. We construct the Georgi-Glashow Lagrangian (also with $\theta $%
-term), and then we define the variation of Lagrangian. We find that the
equation of motion, besides fields, depends on the quantities which are
independent of the representation of the gauge group and a noncommutative
factor. Breaking the gauge symmetry spontaneously gives $q$-dependent masses
of gauge and vacuum manifold. Vacuum manifold is parameterized by the gauge
invariant quantity, which is similar to Seiberg-Witten theory[10] (an
excellent review on this subject is given by Alvarez-Gaume-Hassan[7]), and
for this model, in which the gauge group is quantum group, the parameter of
vacuum manifold is noncommutative. We also derive the field strength
corresponding to the unbroken subgroup and the $q$-dependent BPS bound mass.

The basis of theory presented here are the notion of the differential
calculus on $SU_q(2)$ which was developed by Woronowicz[3] and the $SU_q(2)$
Yang-Mills theory which was constructed by Hirayama[8].

This paper is organized as follows. First we review the $SU_q(2)$ theory in
section II, this review section is taken almost verbatim from Hirayama[8].
In section III we present the $SU_q(n)$ Georgi-Glashow theory. The
discussion of BPS states and the vacuum manifold of the model is presented
in section IV. Section V is devoted for conclusion and outlook.

\section{The $SU_q$(2) and Yang-Mills Theory}

\subsection{$SU_q$(2) Transformation}

We briefly review the $SU_q$(2) theory which was developed by Woronowicz[3].
The fundamental representation of $SU_q$(2) is given by

\begin{equation}
w=\left( 
\begin{array}{cc}
\alpha & -q\gamma ^{*} \\ 
\gamma & \alpha ^{*}
\end{array}
\right) \text{,}  \tag{2.1}
\end{equation}
where $\alpha ,\gamma ,\alpha ^{*},\gamma ^{*}$are operators satisfying
certain algebras[8].

We denote $R$ as polynomial rings which are generated by $I,\alpha ,\gamma
,\alpha ^{*},$and $\gamma ^{*}$and $M_N\left( B\right) $ as a set of $%
N\times N$ matrices whose entries belong to the set $B$. Let $R^{\prime }$
be the set of the representations of $R$ whose operators act on a Hilbert
space $H$[3]. We define the product of $w_1$and $w_2$ as[3]

\begin{equation}
w_1\oplus w_2=\left( 
\begin{array}{cc}
\alpha _2\otimes \alpha _1-q\gamma _2^{*}\otimes \gamma _1 & -q\left( \gamma
_2^{*}\otimes \alpha _1^{*}+\alpha _2\otimes \gamma _1^{*}\right) \\ 
\gamma _2\otimes \alpha _1+\alpha _2^{*}\otimes \gamma _1 & \alpha
_2^{*}\otimes \alpha _1^{*}-q\gamma _2\otimes \gamma _1^{*}
\end{array}
\right) \in M_2\left( R^{\prime }\otimes R^{\prime }\right) ,  \tag{2.2}
\end{equation}
which acts on $H\otimes H$ and it is closely related to the coproduct
defined on $R$. The $*$-operation is the complex conjugate for complex
numbers. The $\otimes $-product of operators $R$ or $R^{\prime }$ is defined
by

\begin{equation}
\left( a_1\otimes a_2\otimes ...\otimes a_n\right) \left( b_1\otimes
b_2\otimes ...\otimes b_n\right) =a_1b_1\otimes a_2b_2\otimes ...\otimes
a_nb_n.  \tag{2.3}
\end{equation}

We denote the set of $w_m\oplus w_{m-1}\oplus ...\oplus w_1$ as $C_m$ and it
has inverse $C_m^{-1}$.

\subsection{Group Theoretic Representation of $w$}

In this subsection we review the group-theoretic representation of $%
SU_q\left( 2\right) $ which was studied by Woronowicz[3]. It turned out that
the representation theory of $SU_q\left( 2\right) $ is quite similar to that
of $SU\left( 2\right) .$ The matrix $W\in M_N\left( R\right) $is said to be
the representation of $w$ if it satisfies[3]

\begin{equation}
\Delta \left( W_{ij}\right) =\left( W\oplus W\right) _{ij}\equiv
\sum_{k=1}^NW_{ik}\otimes W_{kj}\qquad ;i,j=1,2,...,N\quad ,  \tag{2.4}
\end{equation}
where $\Delta $ is the coproduct which is defined as

\begin{equation}
\Delta \left( w\right) \equiv \left( 
\begin{array}{cc}
\Delta \left( \alpha \right) & \Delta \left( -q\gamma ^{*}\right) \\ 
\Delta \left( \gamma \right) & \Delta \left( \alpha ^{*}\right)
\end{array}
\right) =w\oplus w,  \tag{2.5}
\end{equation}

We define the set $C_m^N$ by

\begin{equation}
C_m^N=\left\{ W_m\oplus W_{m-1}\oplus ...\oplus W_1\right\} ,  \tag{2.6}
\end{equation}
where $W_i\in M_N\left( R^{\prime }\right) ,\ i=1,2,...,m$ are the canonical
representations of $w_i\in M_2\left( R^{\prime }\right) $ $,i=1,2,...,m$ ,
respectively.

\subsection{$3D$ Calculus of $SU_q\left( 2\right) $}

The differential calculus of $SU_q\left( 2\right) $ is discussed in
Woronowicz[3], in which coordinates are non-commutative operators. The $3D$
calculus which was studied by Woronowicz[3], is not only left-covariant but
also has simple structure and mysteriously works well even for the higher
order differential calculi. Here we briefly recapitulate the $3D$ calculus
of Woronowicz.

The linear functionals $\chi _0,\chi _1,\chi _2,f_0,f_1,f_2,e$ on $R$ are
defined by

\begin{eqnarray}
\chi _0\left( w\right) &\equiv &\left( 
\begin{array}{cc}
\chi _0\left( \alpha \right) & \chi _0\left( -q\gamma ^{*}\right) \\ 
\chi _0\left( \gamma \right) & \chi _0\left( \alpha ^{*}\right)
\end{array}
\right) =\left( 
\begin{array}{cc}
0 & 1 \\ 
0 & 0
\end{array}
\right) ,  \tag{2.7} \\
\chi _1\left( w\right) &=&\left( 
\begin{array}{cc}
1 & 0 \\ 
0 & -q^2
\end{array}
\right) ,\ \chi _2\left( w\right) =\left( 
\begin{array}{cc}
0 & 0 \\ 
-q & 0
\end{array}
\right) ,\ \chi _k\left( I\right) =0,  \nonumber
\end{eqnarray}

\begin{equation}
f_0\left( w\right) =f_2\left( w\right) =\left( 
\begin{array}{cc}
q^{-1} & 0 \\ 
0 & q
\end{array}
\right) ,\ f_1\left( w\right) =\left( 
\begin{array}{cc}
q^{-2} & 0 \\ 
0 & q^2
\end{array}
\right) ,\ f_k\left( I\right) =1,  \tag{2.8}
\end{equation}

\begin{equation}
e\left( w\right) =\left( 
\begin{array}{cc}
1 & 0 \\ 
0 & 1
\end{array}
\right) ,\ e\left( I\right) =1,\quad k=0,1,2\quad .  \tag{2.9}
\end{equation}

The convolution product of a linear functional $X$ on $R,$ and $a\in R$, is
defined by

\begin{equation}
X*a=\sum_iX\left( a_i^{\prime }\right) a_i^{\prime \prime }\in R  \tag{2.10}
\end{equation}
where $a_i^{\prime }$ and $a_i^{\prime \prime }$ are given by $\Delta \left(
a\right) =\sum_ia_i^{\prime }\otimes $ $a_i^{\prime \prime }$. The
differential operator $d$ is defined by

\begin{equation}
da=\sum_{k=0}^2\left( \chi _k*a\right) \omega _k,\quad a\in R,  \tag{2.11}
\end{equation}
where $\omega _k,\ k=0,1,2,$ are the bases of the space of differential
1-forms. The higher order differential calculus can be defined to maintain
the property

\begin{equation}
d^2=0.  \tag{2.12}
\end{equation}

The Hermitian $\chi _k\left( W\right) ^{\dagger }$ of $\chi _k\left(
W\right) \in M_N\left( C\right) $ is given by

\begin{equation}
\chi _k\left( W\right) ^{\dagger }=\sum_{j=0}^2t_{kj}\chi _j\left( W\right) ,
\tag{2.13}
\end{equation}
where $t_{11}=1,$ $t_{02}=\frac{-1}q,$ and $t_{20}=-q.$

\subsection{Local $SU_q\left( 2\right) $}

Let $x=\left( x^0,x^1,x^2,x^3\right) $ be coordinates of the four
dimensional Minkowski spacetime and $\alpha \left( x\right) ,\gamma \left(
x\right) ,\alpha ^{*}\left( x\right) ,\gamma ^{*}\left( x\right) \in R$ be
the $x$-dependent representations of $\alpha ,\gamma ,\alpha ^{*},\gamma
^{*}\in R$ respectively, as operators acting on the Hilbert space $H$
introduced in 2.1. To discuss the field theory of $SU_q\left( 2\right) $, it
is inevitable to consider the functions of $x,\alpha \left( x\right) ,\gamma
\left( x\right) ,\alpha ^{*}\left( x\right) ,\gamma ^{*}\left( x\right) $
and their derivatives with respect to $x^\mu $. We denote the set of
functions of the form $g\left[ x\right] \equiv g\left( x,\alpha \left(
x\right) ,\gamma \left( x\right) ,\alpha ^{*}\left( x\right) ,\gamma
^{*}\left( x\right) \right) $ by $R^{\prime x}$. The functional $X^x$ on $%
R^{\prime x}$ should be introduced so that $X^x\left( w\left( x\right)
\right) =X\left( w\right) $, e.g.,

\begin{equation}
\chi _k^x\left( w\left( x\right) \right) =\chi _k\left( w\right) ,\
f_k^x\left( w\left( x\right) \right) =f_k\left( w\right) ,\ e^x\left(
w\left( x\right) \right) =e\left( w\right) ,\quad k=0,1,2\ ,  \tag{2.14}
\end{equation}
where $w\left( x\right) $ is defined by

\begin{equation}
w\left( x\right) =\left( 
\begin{array}{cc}
\alpha \left( x\right) & -q\gamma ^{*}\left( x\right) \\ 
\gamma \left( x\right) & \alpha ^{*}\left( x\right)
\end{array}
\right) ,  \tag{2.15}
\end{equation}
and $w,\chi _k,f_k,$ and $e$ are those defined hitherto. Recalling (2.11),
the differential operator $d^x$ should be defined to act on $g\left[
x\right] $ as

\begin{equation}
d^xg\left[ x\right] =\sum_{k=0}^2\left( \chi _k^x*g\left[ x\right] \right)
\omega _k^x+\left( \partial _\mu g\left[ x\right] \right) dx^\mu , 
\tag{2.16}
\end{equation}
where $\omega _k^x$ ,$k=0,1,2$ are the analogue of the previous $\omega _k$
and $\partial _\mu g\left[ x\right] $ is the conventional partial derivative
of $g\left[ x\right] $ with respect to the explicit $x$-dependence of $%
g\left[ x\right] $. A consistent set of rules is derived from the result of
Woronowicz[3] by supposing that $\omega _k^x$ and $d^xg\left[ x\right] $
decompose as $\omega _{k,\mu }^xdx^\mu $ and $\left( D_\mu g\left[ x\right]
\right) dx^\mu $ respectively, and assuming that $\left\{ dx^\mu ,dx^\nu
\right\} =\left[ dx^\mu ,\omega _{k,\nu }^x\right] =\left[ dx^\mu ,a\right]
=0,$ $a\in $ $R^{\prime x},$ $\mu ,\nu =0,1,2,3$. We call the above
procedure as the Z-procedure[8]. The Z-procedure leads us to the following
definition of the partial derivative $D_\mu g\left[ x\right] $ of $g\left[
x\right] \in $ $R^{\prime x}$,

\begin{equation}
D_\mu g\left[ x\right] =\sum_{k=0}^2\left( \chi _k^x*g\left[ x\right]
\right) \omega _{k,\mu }^x+\partial _\mu g\left[ x\right] .  \tag{2.17}
\end{equation}

\subsection{The $SU_q$(2) Yang-Mills Theory}

In this subsection, we briefly review the $SU_q$(2) Yang-Mills theory
constructed by Hirayama[8]. We suppose that the components of the gauge
field, $A_{k,\mu }(x)$ $,\;i=0,1,2,\;\mu =0,1,2,3$ respectively. We
postulate that

\begin{equation}
A_{\mu ,i}(x)W(x)=W(x)f_i(\omega )A_{\mu ,i}(x),  \tag{2.18}
\end{equation}

\begin{equation}
A_{\mu ,i}(x)A_{\nu ,j}(x)=c_{ji}A_{\nu ,j}(x)A_{\mu ,i}(x),  \tag{2.19}
\end{equation}

\begin{equation}
A_{i,\mu }^{\dagger }(x)=\stackrel{2}{\stackunder{k=0}{\sum }}t_{ji}A_{j,\mu
}(x),  \tag{2.20}
\end{equation}
where $t_{02}=-q,t_{20}=-q^{-1},$ and $t_{11}=1$.

Throughout this subsection we denote

\[
W(x)\equiv W_m(x)\oplus W_{m-1}(x)\oplus ...\oplus W_1(x), 
\]

\[
W^{\prime }(x)\equiv W_n^{\prime }(x)\oplus W_{n-1}^{\prime }(x)\oplus
...\oplus W_1^{\prime }(x). 
\]
The vector field $A_{k,\mu }(x)$ transform as

\begin{equation}
A_\mu ^W(x)=W(x)(I_{m-1}\otimes A_\mu (x))W^{-1}(x)-\frac 1{ig}(D_\mu
W(x))W^{-1}(x),  \tag{2.21}
\end{equation}
where

\begin{equation}
A_\mu (x)=\stackrel{2}{\stackunder{k=0}{\sum }}A_{\mu ,k}(x)\chi _k(W_1), 
\tag{2.22}
\end{equation}
the field in the gauge $W(x)$. In (2.21), $g$ is the gauge coupling
constant, $D_\mu W(x)$ is defined by

\begin{equation}
D_\mu W(x)=\stackrel{m}{\stackunder{l=1}{\sum }}\stackrel{}{\stackunder{}{%
W_m(x)\oplus ...\oplus D_\mu W_l(x)\oplus ...\oplus W_1(x),}}  \tag{2.23}
\end{equation}
and $\chi _k(W_1)$ is equal to $\chi _k^x(W_1)$. The gauge transform $(A_\mu
^W(x))^{W^{\prime }}$of $A_\mu ^W(x)$ by $W^{^{\prime }}(x)$ is defined by

\begin{eqnarray}
(A_\mu ^W(x))^{W^{\prime }} &=&(W^{\prime }(x)\otimes I_m)(I_n\otimes A_\mu
^W(x))(W^{\prime -1}(x)\otimes I_m)  \tag{2.24} \\
&&-\frac 1{ig}(D_\mu W^{^{\prime }}(x))W^{^{\prime }-1}(x)\otimes I_m. 
\nonumber
\end{eqnarray}
Then we have

\begin{equation}
(A_\mu ^W(x))^{W^{\prime }}=A_\mu ^{W^{\prime }\oplus W}(x).  \tag{2.25}
\end{equation}

We define the field strength $F_{\mu \nu }^W(x)$ in the gauge $W(x)$ by

\begin{equation}
F_{\mu \nu }^W(x)=\left[ \nabla _\mu ^W,\nabla _\nu ^W\right] ,\nabla _\mu
^W=D_\mu +igA_\mu ^W(x),  \tag{2.26}
\end{equation}
then we find that

\begin{equation}
F_{\mu \nu }^W(x)=W(x)(I_{m-1}\otimes F_{\mu \nu }(x))W^{-1}(x),  \tag{2.27}
\end{equation}

\begin{equation}
F_{\mu \nu }(x)=\stackrel{2}{\stackunder{k=0}{\sum }}F_{k,\mu \nu }(x)\chi
_k(W_1).  \tag{2.28}
\end{equation}
The transformation law of $F_{\mu \nu }^W(x)$ is given by

\begin{eqnarray}
(F_{\mu \nu }^W(x))^{W^{\prime }} &=&(W^{\prime }(x)\otimes I_m)(I_n\otimes
F_{\mu \nu }^W(x))(W^{\prime -1}(x)\otimes I_m)\text{ }  \tag{2.29} \\
&=&F_{\mu \nu }^{W^{\prime }\oplus W}(x)\text{ .}  \nonumber
\end{eqnarray}

The Lagrangian density of the local $SU_q(2)$ invariant field theory should
be independent of the choice of $W(x)$, the dimensionality $N$ and the
integer $m$.

We begin with defining $S_{kl}^W$ by

\begin{equation}
S_{kl}^W=tr(\rho ^N\chi _k(W)(\rho ^N)^2\chi _l(W),W\in C_1^N,k,l=0,1,2, 
\tag{2.30}
\end{equation}
where $\rho ^N$ is given by $\rho ^N=(\sigma ^N)^{-1}.$ If we define $K_N$ by

\begin{equation}
K_N=-\frac q8\left( S_{20}^W\right) ^{-1},W\in C_1^N,  \tag{2.31}
\end{equation}
then the product $K_NS_{kl}^W$ is independent of $W$. Then the gauge
invariant Lagrangian is

\begin{equation}
L_{GG}^W(x)=K_N\text{ }tr(\sigma ^N\tau (W)F^{W,\mu \nu }\tau ^{\dagger
}(W)\tau (W)F_{\mu \nu }^W\tau ^{\dagger }(W)\text{ })\text{ ,}  \tag{2.32}
\end{equation}
where

\begin{equation}
\tau (W)=(W_m(x)\oplus ...\oplus W_2(x)\oplus \rho ^NI)W^{-1}(x)\text{ .} 
\tag{2.33}
\end{equation}

\section{The $SU_q(n)$ Georgi-Glashow Theory}

\subsection{Gauge Field and Scalar Field}

We introduce the gauge field and the scalar field which are fields that
present in the Georgi-Glashow model. We consider the components of gauge
fields and scalar fields are $A_{k,\mu }(x)$ and $\phi _i(x),$where$%
\;i=0,1,...,n^2-2,$ and $\mu =0,1,2,3$ respectively.

If $\Psi _i,i=0,1,...,n^2-2$ are fields, then we generalize the Hirayama's
postulate to

\begin{equation}
\Psi _i(x)W(x)=W(x)f_i(\omega )\Psi _i(x),  \tag{3.1}
\end{equation}

\begin{equation}
\Psi _i(x)\Psi _j(x)=c_{ji}\Psi _j(x)\Psi _i(x).  \tag{3.2}
\end{equation}

From the equation (2.22), we generalize the vector field $A_{k,\mu }(x)$ to

\begin{equation}
A_\mu (x)=\stackrel{n^2-2}{\stackunder{k=0}{\sum }}A_{\mu ,k}(x)\text{ }\chi
_k(W_1),  \tag{3.3}
\end{equation}
and its transformation are given by

\begin{equation}
A_\mu ^W(x)=W(x)(I_{m-1}\otimes A_\mu (x))W^{-1}(x)-\frac 1{ig}(D_\mu
W(x))W^{-1}(x),  \tag{3.4}
\end{equation}
and for the scalar field $\phi (x)$

\begin{equation}
\phi (x)=\stackrel{n^2-2}{\stackunder{k=0}{\sum }}\phi _k(x)\text{ }\chi
_k(W_1),  \tag{3.5}
\end{equation}

\begin{equation}
\phi ^W(x)=W(x)(I_{m-1}\otimes \phi (x))W^{-1}(x).  \tag{3.6}
\end{equation}

The additional property of the gauge field is

\begin{equation}
A_{i,\mu }^{\dagger }(x)=\stackrel{n^2-2}{\stackunder{k=0}{\sum }}t_{ji}%
\text{ }A_{j,\mu }(x)\text{ ,}  \tag{3.7}
\end{equation}

and for the generator is

\begin{equation}
\chi _k\left( W_1\right) ^{\dagger }=\sum_{j=0}^{n^2-2}t_{kj}\text{ }\chi _j%
\text{ }\left( W_1\right) \text{ .}  \tag{3.8}
\end{equation}

If we transform $A_\mu ^W(x)$ to $(A_\mu ^W(x))^{W^{\prime }}$ and $\phi
^W(x)$ to $(\phi ^W(x))^{W^{\prime }}$ , then we get

\begin{eqnarray}
(A_\mu ^W(x))^{W^{\prime }} &=&(W^{\prime }(x)\otimes I_m)(I_n\otimes A_\mu
^W(x))(W^{\prime -1}(x)\otimes I_m)  \tag{3.9} \\
&&-\frac 1{ig}(D_\mu W^{^{\prime }}(x))W^{^{\prime }-1}(x)\otimes I_m, 
\nonumber
\end{eqnarray}

\begin{equation}
(\phi ^W(x))^{W^{\prime }}=(W^{^{\prime }}(x)\otimes I_m)(I_n\otimes \phi
^W(x))(W^{\prime -1}(x)\otimes I_m).  \tag{3.10}
\end{equation}

Thus we have

\begin{equation}
(A_\mu ^W(x))^{W^{\prime }}=A_\mu ^{W^{\prime }\oplus W}(x),  \tag{3.11}
\end{equation}

\begin{equation}
(\phi ^W(x))^{W^{\prime }}=\phi ^{W^{\prime }\oplus W}(x),  \tag{3.12}
\end{equation}
which have the same form as the equation (2.25).

For the scalar part, the transformation law of $\nabla _\mu ^W\phi ^W$ is
also given by

\begin{eqnarray}
(\nabla _\mu ^W\phi ^W(x))^{W^{\prime }} &=&(W^{\prime }(x)\otimes
I_m)(I_n\otimes \nabla _\mu ^W\phi ^W(x))(W^{\prime -1}(x)\otimes I_m)\quad 
\tag{3.13} \\
&=&(\nabla _\mu \phi (x))^{W^{\prime }\oplus W}.  \nonumber
\end{eqnarray}

\subsection{The Construction of Georgi-Glashow Lagrangian}

A similar reason from the previous section can be applied that the
Lagrangian density of the local $SU_q(n)$ invariant field theory should be
independent of the choice of $W(x)$, the dimensionality $N$ and the integer $%
m$.

We begin with defining $S_{kl}^W$ by

\begin{equation}
S_{kl}^W=tr(\rho ^N\chi _k(W)(\rho ^N)^2\chi _l(W),W\in C_1^N,\quad \text{ }%
k,l=0,1,...,n^2-2  \tag{3.14}
\end{equation}
where $\rho ^N$ is given by $\rho ^N=(\sigma ^N)^{-1}.$ If we define $K_N$ by

\begin{equation}
K_N=-\frac q8\left( S_{ij}^W\right) ^{-1},W\in C_1^N,\text{ \qquad }%
i,j=0,1,...,n^2-2  \tag{3.15}
\end{equation}
then the product $K_NS_{kl}^W$ is independent of $W$.

We now define the $\tau $-quantities of any function $F^W(x)$ by

\begin{equation}
(F^W(x))^\tau =\tau (W)F^W(x)\tau ^{\dagger }(W),  \tag{3.16}
\end{equation}
where

\begin{equation}
\tau (W)=(W_m(x)\oplus ...\oplus W_2(x)\oplus \rho ^NI)W^{-1}(x),  \tag{3.17}
\end{equation}
and $F^W(x)$ is function $F(x)$ and transform with respect to the gauge
transformation $W(x)\in C_m^N$. If we transform the $\tau $-quantities of
any function $F^W(x)$ by the gauge transformation $W^{\prime }(x)\in C_n^N,$
then its transform to

\begin{equation}
\left( F^W\left( x\right) \right) ^\tau \rightarrow \left( F^{W^{\prime
}\oplus W}(x)\right) ^\tau =\tau (W^{\prime }\oplus W)F^{W^{\prime }\oplus
W}(x)\tau ^{\dagger }(W^{\prime }\oplus W).  \tag{3.18}
\end{equation}

Then we construct the Georgi-Glashow Lagrangian by

\begin{eqnarray}
L_{GG}^W(x) &=&K_N\text{ }tr\left( 
\begin{array}{c}
\end{array}
\sigma ^N\right( -\frac 14(F_{\mu \nu }^W)^\tau \ \left( F^{W,\mu \nu
}\right) ^\tau  \tag{3.19} \\
&&+(\nabla _\mu ^W\phi ^W)^{\tau ,\dagger }\left( \nabla ^{W,\mu }\phi
^W\right) ^\tau +\lambda \left( \left( \left[ \phi ^{W,\dagger },\phi
^W\right] \right) ^\tau \right) ^2\left) 
\begin{array}{c}
\end{array}
\right)  \nonumber
\end{eqnarray}
where

\begin{eqnarray}
(\nabla _\mu ^W\phi ^W(x))^{\tau ,\dagger } &=&(\tau (W)\nabla _\mu ^W\phi
^W(x)\tau ^{\dagger }(W))^{\dagger }  \tag{3.20} \\
&=&\tau (W)(\nabla _\mu ^W\phi ^W(x))^{\dagger }\tau ^{\dagger }(W) 
\nonumber
\end{eqnarray}

As we expect, the pseudoscalar quantity is also gauge invariant, i.e.,

\begin{equation}
K_N\text{ }tr\left( \sigma ^N(F_{\mu \nu }^W)^\tau (\widetilde{F}^{W,\mu \nu
})^\tau \right) ,  \tag{3.21}
\end{equation}
where $\widetilde{F}^{W,\mu \nu }=\frac 12\varepsilon ^{\mu \nu \rho \sigma
}F_{\rho \sigma }^W$ and $\varepsilon ^{\mu \nu \rho \sigma }$ is the
Levi-Civita symbol for 3+1 dimension.

Then according to Witten[9], we can construct Georgi-Glashow model with an
additional $\theta -$term

\begin{eqnarray}
L_{GG}^W(x) &=&K_N\text{ }tr\left( 
\begin{array}{c}
\end{array}
\sigma ^N\right( -\frac 14(F_{\mu \nu }^W)^\tau (F^{W,\mu \nu })^\tau \  
\tag{3.22} \\
&&+(\nabla _\mu ^W\phi ^W)^{\tau ,\dagger }\left( \nabla ^{W,\mu }\phi
^W\right) ^\tau +\lambda \left( \left( \left[ \phi ^{W,\dagger },\phi
^W\right] \right) ^\tau \right) ^2  \nonumber \\
&&+\frac{\theta e^2}{32\pi ^2}\left( F_{\mu \nu }^W\right) ^\tau \left( 
\widetilde{F}^{W,\mu \nu }\right) ^\tau \left) 
\begin{array}{c}
\end{array}
\right)  \nonumber
\end{eqnarray}
where $\theta $ is a real parameter and $e$ is the charge unit.

\subsection{Equation of Motion}

Before we derive the equation of motion from Lagrangian (3.19) and (3.22),
first we must define the variation of the Lagrangian. Given any function of $%
\Psi _i$ and $D_\mu \Psi _i$, where $\Psi _i$ is a field with $%
i=0,1,...,n^2-2$, then variation of $L(\Psi _i,D_\mu \Psi _i)$ is defined by

\begin{equation}
\delta L(\Psi _i,D_\mu \Psi _i)=\stackunder{i}{\sum }\left( \delta \Psi _i%
\frac{\partial L}{\partial \Psi _i}+\delta \left( D_\mu \Psi _i\right) \frac{%
\partial L}{\partial \left( D_\mu \Psi _i\right) }\right) =0,  \tag{3.23}
\end{equation}
where$\frac \partial {\partial \Psi _i}$ and $\frac \partial {\partial
\left( D_\mu \Psi _i\right) }$ are the usual partial derivative of $\Psi _i$
and $D_\mu \Psi _i$, respectively. From the above definition we get the
equation of motion 
\begin{equation}
\frac{\partial L}{\partial \Psi _i}-D_\mu \left( \frac{\partial L}{\partial
\left( D_\mu \Psi _i\right) }\right) =0.  \tag{3.24}
\end{equation}
We can write the equation (3.19) in component fields

\begin{equation}
L_{GG}=\stackrel{2}{\stackunder{k,l=0}{\sum }}K_NS_{kl}\left( -\frac
14F_{k,\mu \nu }F_l^{\mu \nu }+\stackunder{j}{\sum }t_{jk}(\nabla _\mu \phi
)_j^{\dagger }(\nabla ^\mu \phi )_l+\lambda \left[ \phi ^{\dagger },\phi
\right] _k\left[ \phi ^{\dagger },\phi \right] _l\right) .\   \tag{3.25}
\end{equation}
From equation (3.23), we can define the energy-momentum tensor as

\begin{equation}
T_{\mu \nu }\equiv \stackunder{l}{\sum }\stackrel{n}{\stackunder{(i)=1}{\sum 
}}\left( D_\mu \Psi _l^{(i)}\right) \frac{\partial L}{\partial \left( D^\mu
\Psi _l^{(i)}\right) }-\eta _{\mu \nu }L\text{ },\text{ }(i)=1,2,...n, 
\tag{3.26}
\end{equation}
where index $(i)$ are numbers of fields and $\eta _{\mu \nu }=diag(-1,1,1,1) 
$ is the metric tensor. Then by equation (3.24), we get the equation of
motion for the Lagrangian (3.25)

\begin{equation}
-\frac 12\stackunder{k,l}{\sum }K_N(S_{kl}^W+c_{kl}S_{lk}^W)\left[ \delta
_{ik}D_\nu F_l^{\nu \rho }-ig\stackunder{p}{\sum }d_{kip}A_{p,\nu }F_l^{\nu
\rho }\right] =j_i^\rho ,  \tag{3.27}
\end{equation}
where

\begin{eqnarray}
j_i^\rho &=&ig\sum K_NS_{kl}^W\left( \stackunder{j}{\sum }t_{jk}\right( 
\stackunder{m,n}{\sum }d_{jmn}t_{im}\phi _n^{\dagger }(\nabla ^\rho \phi )_l
\tag{3.28} \\
&&-\stackunder{n}{\sum }d_{lin}\left( (D^\rho \phi _j^{\dagger })c_{ji}-ig%
\stackunder{p,q}{\sum }d_{jpq}c_{qi}c_{pi}\phi _q^{\dagger }A_p^{\rho
,\dagger }\right) \phi _n\left) 
\begin{array}{c}
\end{array}
\right) \   \nonumber
\end{eqnarray}

If we compare with the equation of motion for the Lagrangian (3.22), that is 
\begin{eqnarray}
&&\stackunder{}{\sum_{k,l}}K_N(S_{kl}^W+c_{kl}S_{lk}^W)[-\frac 12(\delta
_{ik}D_\nu F_l^{\nu \rho }-ig\stackunder{p}{\sum }d_{kip}A_{p,\nu }F_l^{\nu
\rho }\quad  \tag{3.29} \\
&&+\frac{\theta e^2}{16\pi ^2}(\delta _{ik}D_\nu \widetilde{F}_l^{\nu \rho
}-ig\stackunder{p}{\sum }d_{kip}A_{p,\nu }\widetilde{F}_l^{\nu \rho })] 
\nonumber \\
&=&j_i^\rho ,  \nonumber
\end{eqnarray}
where $j_i^\rho $ is the same as (3.28), except there is an additional $%
\theta $-term. If we want to preserve Witten's theory[9], i.e., the $\theta
- $term does not affect the equation of motion, then we must impose the
constraint 
\begin{equation}
\stackunder{k,l}{\sum }K_N(S_{kl}^W+c_{kl}S_{lk}^W)(\delta _{ik}D_\nu 
\widetilde{F}_l^{\nu \rho }-ig\stackunder{p}{\sum }d_{kip}A_{p,\nu }%
\widetilde{F}_l^{\nu \rho })=0.  \tag{3.30}
\end{equation}
We called the equation (3.30) as the Bianchi constraint.

In the case of classical Lie group, the Bianchi identity, $\nabla _\mu 
\widetilde{F}^{\mu \nu }=0,$ where $\nabla _\mu =\partial _\mu +igA_\mu $,
and the Bianchi constraint, equation (3.30), coincides. But in the case of
quantum group they are different, because the Bianchi identity comes from
the geometry while the Bianchi constraint comes from the variation of
Lagrangian. The relation between them is not clear until now.

We see from equations (3.25) until (3.30), there are always appear
quantities $K_NS_{kl}^W$ which are always independent of the choice of the
representation of the gauge group and a noncommutative factor $c_{lk}$. We
will see later that both quantities are also appear in the parameter of the
vacuum manifold, i.e. $u_2$, in the field strength corresponding with
unbroken subgroups, and in the BPS bound mass.

\section{BPS States and Vacuum Manifold of the Model}

\subsection{Vacuum Manifold of the Model}

In this section we begin to find the vacuum configuration in this theory. We
start with the energy-momentum tensor of the model which can be derived from
Lagrangian (3.25), that is (without the $\theta $-term) 
\begin{eqnarray}
T_{\mu \nu } &=&\stackunder{p,q,l}{\sum }K_NS_{pq}^W[-\frac 12(\delta
_{lp}+c_{lp}\delta _{lq})(D_\mu A_l^\rho )F_{q,\nu \rho }  \tag{4.1} \\
&&+\stackunder{j}{\sum }t_{jp}((D_\mu \phi )_l(\nabla _\nu \phi )_j^{\dagger
}\delta _{lq}+(D_\mu \phi )_l^{\dagger }(\nabla _\nu \phi )_q\delta
_{lj}]-\eta _{\mu \nu }L_{GG}.  \nonumber
\end{eqnarray}

Then we define a norm denoted by$\left\| \quad \right\| $, i.e., $\left\|
\quad \right\| $ $:F_{nc}\rightarrow {\Bbb R}$, where $F_{nc}$ is a
noncommutative field and ${\Bbb R}$ is a real field , such that 
\begin{equation}
\left\| T_{\mu \nu }\right\| \geq 0,  \tag{4.2}
\end{equation}
and it vanishes only if 
\begin{equation}
F_a^{\mu \nu }=0,\quad \nabla _\mu \phi =0,\quad V(\phi )=0.  \tag{4.3}
\end{equation}

The first equation in (4.3) implies that in the vacuum, $F_a^{\mu \nu }$ is
pure gauge and the last two equations define the Higgs vacuum. The structure
of the space of vacua is determined by 
\begin{equation}
V(\phi )=\stackunder{k,l}{\sum }K_NS_{kl}^W\left[ \phi ^{\dagger },\phi
\right] _k\left[ \phi ^{\dagger },\phi \right] _l=0.  \tag{4.4}
\end{equation}
Therefore, the Higgs vacuum is defined by $\left[ \phi ^{\dagger },\phi
\right] =0$, which implies that $\phi $ takes values in the Cartan
subalgebra of the gauge group $SU_q(n)$. We denote by $U_q(1)^{n-1}$ a
subgroup of $SU_q(n)$, which is generated by elements of the Cartan
subalgebra of the gauge group $SU_q(n)$. It is clear that $U_q(1)^{n-1}$ is
the unbroken subgroup of $SU_q(n)$ which keeps the Higgs vacuum invariant.

There the Georgi-Glashow model has a family of vacuum states. Vacuum
manifold, which is formed by the potential (4.4), parameterized by gauge
invariant quantities. For this model, we have the gauge invariant quantity
parameterizing the space of vacua, that is

\begin{equation}
u_n=K_N\text{ }tr(\sigma ^N\left( \left[ \phi ^W\right] ^\tau \right) ^n)%
\text{ ,}  \tag{4.5}
\end{equation}
which is similar to Seiberg-Witten theory [10]. For $SU_q(2)$ gauge group,
the parameter $u_n$ in (4.5) is

\begin{equation}
u_2=K_{N\text{ }}tr(\sigma ^N\left( \left[ \phi ^W\right] ^\tau \right) ^2)%
\text{ .}  \tag{4.6}
\end{equation}
As we mention above, if we write the above equations in their components,
then quantities $K_NS_{kl}$ appears in the parameter of the vacuum manifold.
Then, up to a gauge transformation , we can take $\phi =a\chi _1$, so the
parameter $u_2$ in (4.6) becomes

\begin{equation}
u_2=\frac 18\left( 1+q^2\right) a^2\equiv u,  \tag{4.7}
\end{equation}
where $a\in {\Bbb C}_{nc}$ and ${\Bbb C}_{nc}$ is the noncommutative complex
field.

If we take values of $\phi $ in Cartan subalgebra, i.e., $\phi =a\chi _1$,
then the Lagrangian (3.25) becomes 
\begin{eqnarray}
L_{GG} &=&-2qK_NS_{02}^w\left[ A_{2,\mu }^{*}(D_\nu D^\nu )A_2^\mu +\frac
12A_{2,\mu }^{*}(q^{-6}\left( 1+q^2\right) )^2a^{*}aA_2^\mu \right] 
\tag{4.8} \\
&&-2q^{-1}K_NS_{20}^W\left[ A_{0,\mu }^{*}(D_\nu D^\nu )A_0^\mu +\frac
12A_{0,\mu }^{*}(q^6\left( 1+q^2\right) )^2a^{*}aA_0^\mu \right]  \nonumber
\\
&&+2K_NS_{11}^WA_{1,\mu }^{*}(D_\nu D^\nu )A_1^\mu +...  \nonumber
\end{eqnarray}
where dots denote higher order terms. From the above Lagrangian we can read
off the masses of the gauge fields as follow

\begin{eqnarray}
m_0 &=&q^6\left( 1+q^2\right) \left\| a^{*}a\right\| ^{1/2}  \tag{4.9} \\
m_1 &=&0  \nonumber \\
m_2 &=&q^{-6}\left( 1+q^2\right) \left\| a^{*}a\right\| ^{1/2}.  \nonumber
\end{eqnarray}

\subsection{$U_q(1)^{n-1}$-Field Strength and BPS States}

\subsubsection{$U_q(1)^{n-1}$-Field Strength}

In this subsection, we derive the solution of the second equation in (4.3),
then find the field strength corresponding to the unbroken part of the gauge
group, i.e., $U_q(1)^{n-1}$.

Let $\phi _i^{(v)}$denote the field $\phi _i$ in a Higgs vacuum. It then
satisfies the equations

\begin{eqnarray}
\left[ \phi ^{(v)\dagger },\phi ^{(v)}\right] &=&0,  \tag{4.10} \\
D_\mu \phi ^{(v)}+ig\left[ A_\mu ,\phi ^{(v)}\right] &=&0.  \nonumber
\end{eqnarray}
We find that the solution of the second equation in (4.9) is

\begin{equation}
A_{j,\mu }=\frac 1{ig}\left( -\stackunder{p,q,i}{\sum }(M^{-1})_{jp}d_{pqi}%
\phi _q^{(v)\dagger }(D_\mu \phi _i^{\left( v\right) }+\stackunder{p,k}{\sum 
}(M^{-1})_{jp}M_{pk}^{\prime }\phi _k^{(v)\dagger }A_\mu \right) , 
\tag{4.11}
\end{equation}
where

\begin{equation}
M_{pj}=\stackunder{q,i,k}{\sum }d_{pqi}d_{ijk}c_{kj}\phi _q^{(v)\dagger
}\phi _k^{(v)},  \tag{4.12}
\end{equation}

\begin{equation}
M_{pj}^{\prime }=\stackunder{q,i,j,l}{\sum }t_{kj}d_{pqi}d_{ijl}c_{lk}\phi
_q^{(v)\dagger }\phi _l^{(v)},  \tag{4.13}
\end{equation}
and $M^{-1}$ is the inverse of $M$ with $det_fM=\stackunder{\sigma }{\sum }%
(-f(c_{lk}))^{l(\sigma )}M_{\sigma (1)}^1...M_{\sigma (3)}^3$, where $%
l(\sigma )$ is the minimal number of inversions in permutation $\sigma $ [1].

If we define 
\begin{equation}
h_{j,\mu }(\phi ^{(v)},\phi ^{(v)\dagger },D_\mu \phi ^{(v)})\equiv 
\stackunder{p,q,i}{\sum }(M^{-1})_{jp}d_{pqi}\phi _q^{(v)\dagger }D_\mu \phi
_i^{(v)},  \tag{4.14}
\end{equation}
\begin{equation}
\widetilde{g}_j(\phi ^{(v)},\phi ^{(v)\dagger })\equiv \stackunder{p,k}{\sum 
}(M^{-1})_{jp}M_{pk}^{\prime }\phi _k^{(v)\dagger },  \tag{4.15}
\end{equation}
then we get 
\[
f_{j,\mu \nu }=-\frac 1{ig}(D_\mu h_{j,\nu }-D_\nu h_{j,\mu })+\frac 1{ig}%
\stackunder{k,l}{\sum }d_{jkl}h_{k,\mu }h_{l,\nu }+\frac 1{ig}\widetilde{g}%
_j(D_\mu A_\nu -D_\nu A_\mu ), 
\]
with constraints 
\begin{equation}
\stackunder{k,l}{\sum }d_{jkl}\widetilde{g}_kA_\mu \widetilde{g}_lA_\nu =0, 
\tag{4.16}
\end{equation}
\begin{equation}
D_\mu \widetilde{g}_j+\frac 1{ig}\stackunder{k,l}{\sum }d_{jkl}h_{k,\mu
}g_l=0.  \tag{4.17}
\end{equation}

The field strength $F_{\mu \nu }$ corresponding to the unbroken part of $%
SU_q(n)$ can be identified as 
\begin{eqnarray}
F_{\mu \nu } &=&K_Ntr\left( \sigma ^N\left( \phi ^{\left( v\right) W}\right)
^\tau \ \left( F_{\mu \nu }^W\right) ^\tau \right)  \tag{4.18} \\
&=&\stackunder{k,l}{\sum }K_NS_{kl}^W\phi _k^{(v)}F_{l,\mu \nu }\text{.} 
\nonumber
\end{eqnarray}

\subsubsection{BPS\ States}

In this subsection, we derive the Bogomol'nyi bound[14] on the mass of dyon
in terms of its electric and magnetic charge, which are sources for the
equation (4.18). We define the electric and magnetic charge as 
\begin{equation}
q\equiv \stackunder{\partial X}{\oint }E_adS^a=\stackunder{X}{\int }\nabla
_aE^ad^3x\text{,}  \tag{4.19}
\end{equation}
\begin{equation}
g\equiv \stackunder{\partial X}{\oint }B_adS^a=\stackunder{X}{\int }\nabla
_aB^ad^3x\text{, \quad }a=1,2,3\text{,}  \tag{4.20}
\end{equation}
respectively, where $X$ is a manifold and $\partial X$ is the boundary of $X$%
. From equation (4.18) and the Bianchi identity, $\nabla _\mu \widetilde{F}%
^{\mu \nu }=0$, the electric and magnetic charge can be written as 
\begin{equation}
q=\stackunder{k,l}{\sum }K_NS_{kl}^W\int \left[ \left( \nabla _a\phi
_k\right) E_l^a+\phi _k\left( \nabla _aE_l^a\right) \right] d^3x\text{,} 
\tag{4.21}
\end{equation}
\begin{equation}
g=\stackunder{k,l}{\sum }K_NS_{kl}^W\int \left( \nabla _a\phi _k\right)
B_l^ad^3x\text{,}  \tag{4.22}
\end{equation}
where $E_l^a=F_l^{0a}$ and $B_l^a=-\frac 12\varepsilon ^{abc}F_{bc,l}$, $%
a,b,c=1,2,3$.

Now the dyon mass is given by 
\begin{eqnarray}
M &\equiv &\left\| \int T_{00}d^3x\right\| \geq \left\| \int
L_{GG}d^3x\right\|  \tag{4.23} \\
&\geq &\left\| \stackunder{k,l}{\sum }K_NS_{kl}^W\right[ \sqrt{2}\int \left( 
\stackunder{j}{\sum }t_{jk}\left( \nabla _a\phi _j\right) ^{\dagger
}E_l^a+c_{lk}\left( \nabla _a\phi _l\right) E_k^a\right) d^3x\sin \theta 
\nonumber \\
&&+\sqrt{2}\int \left( \stackunder{j}{\sum }t_{jk}\left( \nabla _a\phi
_j\right) ^{\dagger }B_l^a+c_{lk}\left( \nabla _a\phi _l\right) B_k^a\right)
d^3x\cos \theta \;\ \left] 
\begin{array}{c}
\end{array}
\right\|  \nonumber \\
&\geq &\left\| \stackunder{k,l}{\sum }K_NS_{lk}^W\left( \sqrt{2}c_{kl}\int
\left[ \left( \nabla _a\phi _k\right) E_l^ad^3x\sin \theta +\left( \nabla
_a\phi _k\right) B_l^ad^3x\cos \theta \right] \right) \right\| .  \nonumber
\end{eqnarray}
We see that there exist $U$ transformation such that $\left( S^W\right)
^T=S^WU$, where $T$ denotes transpose of a matrix. Using this, the equation
(4.23) can be written as 
\begin{equation}
M\geq \left\| \stackunder{k,l}{\sum }K_NU_{jl}S_{kj}^W\sqrt{2}c_{lk}\left[
\int \left( \nabla _a\phi _k\right) E_l^ad^3x\sin \theta +\int \left( \nabla
_a\phi _k\right) B_l^ad^3x\cos \theta \right] \right\| .  \tag{4.24}
\end{equation}

We propose that there exists $\epsilon >0$ such that the above equation
becomes 
\begin{equation}
M\geq \frac 1\epsilon \left\| U\right\| \left\| c\right\| \left\| \sqrt{2}%
\left( q\sin \theta +g\cos \theta \right) \right\| ,  \tag{4.25}
\end{equation}
and it turns out that $\epsilon =\left\| U\right\| _{q=1}\left\| c\right\|
_{q=1}$ because the model will reduce to the $SU(2)$ case when $n=2$ and $%
q=1 $[5,6,7,11,12].

\section{Conclusion and Outlook}

In this paper, we have constructed the $SU_q(n)$ Georgi-Glashow model (also
with $\theta $-term). The equation of motion, besides the fields, depends on
quantities $K_N\ S_{kl}^W$ which are independent of the representation of
the gauge group and a noncommutative factor $c_{lk}$. In the case of
classical Lie group, the Bianchi identity and the Bianchi constraint,
equation (3.30), coincides. But in the case of quantum group they are
different, because the Bianchi identity comes from the geometry while the
Bianchi constraint comes from the variation of Lagrangian. The relation
between them is not clear until now. We break the gauge symmetry
spontaneously and this gives rise to the masses of gauge field which depend
on $q$ and $a$, where $a$ is the vacuum parameter. The vacuum manifold is
parameterized by the gauge invariant quantity which depends on a scalar
field $\phi $. We get the field strength corresponding to the unbroken
subgroup, $U_q(1)^{n-1}$, and the $q$-dependent BPS bound mass.

For further work, we extend the problem to the supersymmetric case,
especially the Seiberg-Witten theory[13].

\section{Acknowledgement}

One of us (F.Z) would like to thank YANBINBANG SDM-IPTEK (Habibie
Foundation) for financial support. Work of F.Z and B.G is supported by Hibah
Bersaing VII/2, 1999-2000 project of DIKTI, Minister of Education and
Culture of the Republic of Indonesia. F.Z, B.G and R.M thank the Abdus Salam
International Centre for Theoretical Physics for their warm hospitality
during our visit, where we initiated this project. We also thank P.Silaban
for his encouragement.

\section{References}

1. V. G. Drinfel'd, in {\it Proceedings of the International Congress of Math%
}

\quad (Berkeley, CA, USA, 1986), p.286.

2. M. Jimbo, Lett. Math. Phys. {\bf 10} (1985), p.63.

3. S. L. Woronowicz, Publ. RIMS, Kyoto Univ. {\bf 23} (1987), p.117.

4. P. Podles and S. L. Woronowicz, Commun. Math. Phys. {\bf 130} (1990),

\quad p.381,

\quad U. Carow-Watamura, M. Schlieker, M.Scholl, and S. Watamura,

\quad Int. J. Mod. Phys. {\bf 6} (1991), p.3081.

5. G. 't Hooft, Nuc. Phys. {\bf B79} (1974) p.276, A. M. Polyakov,

\quad JETP Lett. {\bf 20} (1974), p.194.

6. B. Julia and A. Zee, Phys. Rev. {\bf D11} (1975) p.2227.

7. L. Alvarez-Gaume and S. F. Hassan, CERN-TH/96-371.

8. M. Hirayama, Prog. Theor. Phys. {\bf 88} (1992) p.111.

9. E. Witten, Phys. Lett. {\bf 86B} (1979) p.283.

10. N. Seiberg and E. Witten, Nuc. Phys. {\bf B426} (1994) p.19,

\quad \ N. Seiberg and E. Witten, Nuc. Phys. {\bf B431} (1994) p.484.

11. E. B. Bogomol'nyi, Sov. J. Nuc. Phys. {\bf 24} (1976) p.449.

12. M. K. Prasad and C. M. Sommerfield, Phys. Rev. Lett. {\bf 35}

\quad \ (1975) p.760.

13. F. P. Zen, B. E. Gunara, D. P. Hutasoit, R. Muhamad, in progress.

\end{document}